\documentclass[11pt]{article}

\usepackage{amssymb}
\usepackage{graphicx}
\begin{document}

\begin{center}
{\Large\bf Constraints on the Generalized Chaplygin Gas Model from Gamma-Ray Bursts}
\medskip

{ \hspace{2cm} R. C. Freitas$^{a,}$\footnote{e-mail: rc\_freitas@terra.com.br}, S. V. B. Gon\c{c}alves$^{a,}$\footnote{e-mail: sergio.vitorino@pq.cnpq.br} and H. E. S. Velten$^{a,b,}$\footnote{e-mail: velten@cce.ufes.br}, \medskip
\newline \it a Grupo de Gravita\c c\~ao e Cosmologia, Departamento de F\'isica, \\ Universidade Federal do Esp\'{\i}rito Santo,
29075-910, Vit\'oria, Esp\'{\i}rito Santo, Brazil}
\newline {\it b Fakult\"at f\"ur Physik, Universit\"at Bielefeld, Bielefeld 33615, Germany.}
\medskip\end{center}

\begin{abstract}
We study the generalized Chaplygin gas model (GCGM) using Gamma-ray bursts as cosmological probes. In order to avoid the so-called circularity problem we use cosmology-independent data set and Bayesian statistics to impose constraints on the model parameters. We observe that a negative value for the parameter $\alpha$ is favoured in a flat Universe and the estimated value of the parameter $H_{0}$ is lower than that found in literature.       
\end{abstract}

\section{Introduction}
\par
One of the most important problems of Modern Cosmology is the determination of the matter content of the Universe. The rotation curve of spiral galaxies \cite{rotation}, the dynamics of galaxy clusters \cite{dynamics} and structure formation \cite{struc}, indicate that there is about ten times more pressureless matter in the Universe than can be afforded by the baryonic matter. The nature of this dark matter component remains unknown. Moreover, the Type Ia supernovae (SNe Ia) data indicates that the Universe is accelerating \cite{super}. Models considering matter content dominated by an exotic fluid whose pressure is negative \cite{press}, modified gravity theories such as $f(R)$ \cite{mod} and the evolution of an inhomogeneous Universe model described in terms of spatially averaged scalar variables with matter and backreaction source terms \cite{back} are some of the proposals to explain this current phase of the Universe. At the same time, the position of the first acoustic peak in the spectrum of CMB anisotropies, as obtained by WMAP, favours a spatially flat Universe\cite{WM5}. Combining all these data and if we consider the matter content of the Universe dominated by a fluid with negative pressure we have a scenario with a proportion of $\Omega_{m} \sim 0.27$ and $\Omega_{de} \sim 0.73$, with respect to the critical density, for the fractions of the pressureless matter and dark energy, respectively. This scenario is usually called as the concordance cosmological model.
\par
The question is to know what is the nature of the dark matter and dark energy components. For dark matter many candidates have been suggested such as axions, a particle until now undetected which would be a relic of a phase where the grand unified theory was valid \cite{axion}, the lightest supersymmetric particle (LSP) like neutralinos \cite{salati1} and the Kaluza-Klein particles \cite{salati2} that are stable viable Weakly Interacting Massive Particles (WIMPs) and arise in two frameworks: In Universal Extra Dimensions \cite{UED} and in some warped geometries like Randall-Sundrum \cite{RS}. For the dark energy, in the hydrodynamical representations of matter, the most natural candidate is a cosmological constant, but there is a discrepancy of some $120$ orders of magnitude between its theoretical and observed values \cite{cc}. For this reason, other candidates have been suggested like quintessence models that involve canonical kinetic terms of the self-interacting scalar field with the sound speed $c_s^2 = 1$ \cite{quinte} and k-essence models that employ rather exotic scalar fields with non-canonical (non-linear) kinetic terms which typically lead to a negative pressure \cite{kesse}. More recently, a string-inspired fluid has been evoked: The Chaplygin gas \cite{chapl}, that appears as a promising candidate for the dark sector of the Universe. 
\par
The Chaplygin gas is represented by the equation of state
\begin{equation}
p_c = - \frac{A}{\rho_c} \quad ,
\end{equation}
where $p_c$ represents the pressure, $\rho_c$ the fluid density and $A$ is a parameter connected with the sound speed. This equation of state is suggested by a brane configuration in the context of string theories \cite{string}. However, a more general equation of state has been suggested  \cite{gcg}:
\begin{equation}
\label{chapp}
p_c = - \frac{A}{\rho^{\alpha}_c} \quad ,
\end{equation}
where again $p_c$ and $\rho_c$ stand for the generalized Chaplygin gas component and $\alpha$ is a new parameter, which takes the value $1$ for the traditional Chaplygin gas but values larger than $1$, or even negative may be considered. This is the so-called generalized Chaplygin gas.
\par
Much observational data that has been used for comparison with the theoretical cosmological models like the generalized Chaplygin gas model (GCGM). The spectra of anisotropy of cosmic microwave background radiation \cite{berto1}, baryonic acoustic oscillations \cite{BAOcha}, the integrated Sachs-Wolfe effect \cite{SW}, the matter power spectrum \cite{mass}, gravitational lenses \cite {lens}, X-ray data \cite{raiox} and ages estimates of high-$z$ objects \cite{ages} have been used in this sense. Also, constraints from combined data sources have been obtained in \cite{combcha}. Another tool used to make this comparison is the Hubble diagram, the plot of redshift $z$ versus luminosity distance $d_L = \sqrt{\mathcal{L}/4\pi\mathcal{F}}$, where $\mathcal{L}$ is the luminosity (the energy per time produced by the source in its rest frame) and $\mathcal{F}$ is the measured flux, i.e., the energy per time per area measured by a detector. Normally, the SNe Ia data are considered good standard candles and are used to construct the Hubble diagram, because their luminosity are well known \cite{super, SN2}. In particular, constraints on the Generalized Chaplygin gas have been studied in \cite{SNe}. These assumptions rest on a foundation of photometric and spectroscopic similarities between high- and low-redshift SNe Ia. But this discussion is not yet finished \cite{SN3}. The other problem comes from the fact that there still does not exist SNe Ia data with $z > 1.8$. To know the properties and behavior of dark energy for high values of $z$ we will have to wait for new data of the SNe Ia or to find other distance indicators.  In this sense, to extend the comparison between observational data and theoretical models at very high redshift we propose to use Gamma-ray bursts (GRBs) due to the fact that they occur in the range of high z beyond the SNe data found today \cite{Bromm}.

The GRBs are jets that release $\sim 10^{51} - 10^{53}$ ergs or more for a few seconds and becomes, in this brief period of time, the most bright object in the Universe. They were discovered in the sixties by the Vela satellites in the ``Outer Space Treaty" that monitored nuclear explosions in space \cite{hist}. Launched in 1991 The Burst and Transient Source Experiment on the Compton Gamma-Ray Observatory (BATSE on the Compton GRO) \cite{Costa} observations concluded that the angular distribution of the GRBs on the sky is isotropic within statistical limits. This study ruled out the idea that the GRBs are galactic objects, but it is consistent with the bursts being extra-galactic sources at cosmological distances. More recently, the SWIFT mission (launched in 2004) has provided the most accurate GRB data, available in the Swift BAT Catalog.
\par
The search for a self-consistent method to use the GRBs in cosmological problems is intense and promising. But the possibility of using GRBs as standard candles is not a simple question. GRBs are known to have several light curves and spectral properties from which the luminosity of the burst can be calculated once calibrated, and these can make GRBs into standard candles. Just as with SNe Ia, the idea is to measure the luminosity indicators, deduce the source luminosity, measure the observed flux and then use the inverse-square law to derive the luminosity distance. The difficulty arises when these indicators are a priori established through some cosmological model like the concordance one. This means that the parameters of the calibrated relations of luminosity/energy are still coupled to the cosmological parameters derived from a given cosmological model. This is the so called circularity problem. This problem appears in several works that have made use of these GRBs luminosity indicators as standard candles at very high redshift \cite{circular}. It is possible to treat the circularity problem with a statistical approach \cite{statist}. On the other hand, many papers have dealt with the use of so called Amati relation, or the Ghirlanda relation for this purpose \cite{firmani}. However, as argued recently in \cite{petrosian}, these procedure involve many unjustified assumptions which if not true could invalidate the results. In particular, many evolutionary effects can affect the final outcome. However, recently Liang {\it et al.} \cite{liang, liang2010, liang11} made a study considering SNe Ia as first-order standard candles for calibrating GRBs, the second-order standard candles. The sample in reference \cite{liang} was calibrated from the 192 supernovae obtained in \cite{davies}. The updated sample used in \cite{liang2010, liang11} has been obtained and calibrated cosmology-independently from the Union2 (557 data points) compilation \cite{Union2} released by the Supernova Cosmology Project Collaboration. In these articles the authors found relevant constraints on the Cardassian and Chaplygin gas model by adding to the GRB data the SNe Ia (Union2), the Shift parameter of the Cosmic Microwave Background radiation from the seven-year Wilkinson Microwave Anisotropy Probe and the baryonic acoustic oscillation from the spectroscopic Sloan Digital Sky Survey Data Release galaxy sample. The sample obtained in \cite{liang2010} will be used in our analysis. These authors obtain the distance moduli $\mu$ of GRB in the redshift range of SNe Ia and extend this result to very high redshift GRB ($z > 1.4$) in a completely cosmological model-independent way. This approach has been also studied in \cite{calibration}.

Some analysis have been made with the GCGM and the GRBs as distant markers \cite{GRB}. In the reference \cite{bertolami} the authors build a specific distribution of GRB to probe the flat GCGM and the XCDM model. While the GCGM has an equation of state given by expression (\ref{chapp}) the XCDM model is considered in terms of a constant equation of state $\omega = p/\rho < 0$. The main conclusion of this article is that the use of GRBs as a dark energy probe is more limited when compared to SNe Ia. We anticipate that we shall arrive at a similar conclusion. Moreover the XCDM model is better constrained than the GCGM. On the other hand, in \cite{herman} the GCGM and the $\Lambda$CDM model are compared by using the GRB and SNe Ia data to build the Hubble diagram. These authors show through the statistical analysis that the Chaplygin gas model (they use $\alpha = 1$) have the best fit when compared with the data. Also they verify that the transition redshift between the decelerated and the accelerated state of the Universe occurs at $z \sim 2.5 - 3.5$ rather than $z \sim 0.5 - 1$ based on the analysis made with the SNe Ia. Here, for our purpose, we will assume the plausible assumption that GRBs are standard candles and we will use the data from Liang {\it et al.} \cite{liang2010}, calibrated
cosmology-independently from the Union2 compilation of SNe Ia, to constraint the cosmological parameters of the GCGM. We want to show how GRB data could constraint different Chaplygin cosmologies.
\par
This paper is organized as follows. In next section, we described a brief review of GCGM. In section $3$ the luminosity distance $d_L$ is obtained for the GCGM and compared with the observational data. Finally, in section $4$ we present our discussion and conclusions.


\section{The Generalized Chaplygin Gas Model}
\label{sectionCGM}
\par
We consider here an homogeneous and isotropic Universe described by the Friedmann's equation
\begin{equation}
\label{be1}
\biggl(\frac{\dot{a}}{a}\biggr)^2 +  \frac{k}{a^2} = \frac{8\pi G}{3} (\rho_m + \rho_c)\quad,
\end{equation}
where the density $\rho$ has the subscripts $m$ for the matter pressureless fluid and $c$ for the generalized Chaplygin gas with equations of state $p_m = 0$ and $p_c=-A/\rho_{c}^{\alpha}$, respectively. Dot means derivative with respect to the cosmic time $t$. Flat, closed and open spatial sections correspond to $k = 0, 1, -1$ for the constant of the curvature.
\par
In our case, each fluid obeys separately the energy conservation law. The equations and the respective solutions are given by
\begin{eqnarray}
\label{mat}
\dot\rho_m + 3\frac{\dot a}{a}\rho_m &=& 0\quad\rightarrow\quad\quad\rho_m = \frac{\rho_{m0}}{a^3}\quad,\\
\label{chap}
\dot\rho_c + 3\frac{\dot a}{a}\biggl(\rho_c - \frac{A}{\rho^{\alpha}_c}\biggr) &=& 0\quad\rightarrow\quad\quad\rho_c = \rho_{c0}\biggl(\bar{A} + \frac{1 - \bar{A}}{a^{3(1+\alpha)}}\biggr)^{1/(1+\alpha)}\quad,
\end{eqnarray}
where $\rho_{m0} = \rho_m(a_0)$, $\rho_{c0} = \rho_c(a_0) = (A + B)^{1/(1 + \alpha)}$ with $a(t = 0) = a_0 = 1$ being the scale factor today. The new definition of the constant $A$ is given by $\bar{A} = A/\rho_{c0}^{1+\alpha}$ and it is connected to the sound velocity today in the gas by the expression $v_{s_0} = \sqrt{\partial p_c/\partial \rho_c} \Big|_{t_0} = \sqrt{\alpha\bar{A}}$.
\par
Initially, the GCGM behaves like a dust fluid, with $\rho\propto a^{-3} $, while at late times the GCGM behaves as a cosmological constant term, $\rho \propto A^{1/(1 + \alpha)}$. Hence, the GCGM interpolates a matter dominated phase (where the formation of structure occurs) and a de Sitter phase. At the same time, the pressure is negative while the sound velocity is positive, avoiding instability problems at small scales \cite{jerome}.
\par
In order to proceed with data comparison we need to calculate the luminosity distance in the GCGM. Using the expression for the propagation of light and the Friedmann's equation (\ref{be1}), we can express the luminosity distance as 
\begin{equation}
d_L = \frac{a_0^2}{a}r_1 = (1 + z)S[f(z)] \quad ,
\end{equation}
where $r_1$ is the co-moving coordinate of the source and
\begin{eqnarray}
S(x) &=& x\quad\mbox{for}\quad (k = 0) \nonumber\quad ,\\
S(x) &=& \sin x\quad\mbox{for}\quad(k = 1) \quad,\nonumber\\
S(x) &=& \sinh x\quad\mbox{for}\quad(k = - 1)\quad.
\end{eqnarray}
The function $f(z)$ is given by
\begin{equation}
f(z) = \frac{1}{H_0}\int_0^z \frac{dz'}{\{\Omega_{m}(z^{\prime}+ 1)^3 + \Omega_{c}[\bar A + (1 - \bar A)(z^{\prime}+ 1)^{3(1+\alpha)}]^{1/(1+\alpha)} - \Omega_{k}(z^{\prime}+ 1)^2\}^{1/2}} \quad ,
\end{equation}
with the definitions 
\begin{equation}
\Omega_{m} = \frac{8\pi G}{3}\frac{\rho_{m0}}{H_0^2} \quad , \quad
\Omega_{c} = \frac{8\pi G}{3}\frac{\rho_{c0}}{H_0^2} \quad , \quad
\Omega_{k} = - \frac{k}{H_0^2} \quad ,
\end{equation}
and $\Omega_{m0} +\Omega_{c0} + \Omega_{k} = 1$. The final equations have been also expressed in terms of the redshift $z = - 1 + \frac{1}{a}$.
\par
In our numerical calculations we relax the restriction that the pressureless matter component is entirely given
by baryons. We consider the nucleosynthesis results for the baryonic component of the Universe and assume the total pressureless matter density as $\Omega_{m}=\Omega_{b}+\Omega_{dm}$, where $\Omega_{b}h^{2}=0.0223$ and $H_0=100h Km s^{-1} Mpc^{-1}$.  Then, in our notation $\Omega_{dm}$ means the extra dark matter contribution.


\section{The Numerical Results}
\label{sectionNA}
\par
The observational data set used in this article is composed by 42 GRBs from \cite{liang2010, liang11}. As told in the introduction, this sample has been obtained and calibrated cosmology-independently from the Union2 compilation. This fact is of crucial importance to admit GRBs as cosmological probes since the circularity problem described above is avoided. At the same time, this data set allow us to analyse the free parameters of the GCGM for a redshift range larger than the available data from SNIa reaching up to z $\approx$ 6.
It is important to emphasize that with this sample the authors of \cite{liang2010, liang11} have obtained stark constraints on the Cardassian and Chaplygin gas model by combining the GRB data with other cosmological probes.

If we want to have a reliable sample of GRBs to make our analysis, the Hubble diagram for the GRBs should be calibrated from the SN at $z \leq 1.4$. This allows to obtain the following luminosity/energy relations: The $\tau_{lag}-L$ relation, the $V-L$ relation, the $L-E_p$ relation, the $E_{\gamma}-E_p$ relation, the $\tau_{RT}-L$ relation, the $E_{iso}-E_p$ relation, and the $E_{iso}-E_p-t_b$ relation. In general these relations can be written as $\mbox{log}(y) = a + b~\mbox{log}(x)$ (two-variable relations) and $\mbox{log}(y) = a + b_1~\mbox{log}(x_1) + b_2~\mbox{log}(x_2)$ (multi-variable relation). In this relations $y$ is the luminosity in units of erg s$^{-1}$ or energy in units of erg and $x$ is the GRB parameter measured in the rest frame; in the latter expression $x_1$ and $x_2$ are $E_p (1 + z)/(300~\mbox{keV})$ and $t_b/(1 + z)/(1~\mbox{day})$ respectively, and $b_1$ and $b_2$ are the slopes of $x_1$ and $x_2$ respectively. The calibration's process is achieved using two methods: linear interpolation (the bisector of the two ordinary least-squares) and the cubic interpolation (the multiple variable regression analysis). The variables $a$ and $b_i$ are determinated with $1-\sigma$ uncertainties. With the linear interpolation, the error of the interpolated distance modulus can be calculated by $\sigma_{\mu} = ([(z_{i + 1} - z)/(z_{i + 1} - z_i)]^2\epsilon^2_{\mu , i} + [(z - z_i)/(z_{i + 1} - z_i)]^2\epsilon^e_{\mu , i + 1})^{1/2}$, where $\mu$ is the interpolated distance modulus of a source at redshift $z$, $\epsilon^2_{\mu , i}$ and $\epsilon^e_{\mu , i + 1}$ are errors of the SNe, $\mu_i$ and $\mu_{i + 1}$ are the distance moduli of the SNe at nearby redshifts $z_i$ and $z_{i + 1}$, respectively. In the case of the cubic interpolation method the error can be estimated by the expression
$\sigma_{\mu} = (A_0^2\epsilon^2_{\mu , i} + A_1^2\epsilon^2_{\mu , i + 1} + A_2^2\epsilon^2_{\mu , i + 2} + A_3^2\epsilon^2_{\mu , i + 3})^{1/2}$, where
$\epsilon_{\mu , i + j}$ are errors of the SNe and $\mu_{i + j}$ are the distance moduli of the SNe at  nearby redshifts $z_{i + j}$ (index $j$ run from $0$ to $3$) with:
\begin{eqnarray}
A_0 &=&\frac{ [(z_{i + 1} - z)(z_{i + 2} - z)(z_{i + 3} - z)]}{[(z_{i + 1} - z_i)(z_{i + 2} - z_i)(z_{i + 3} - z_i)]}\quad;\nonumber\\
A_1 &=& \frac{[(z_{i} - z)(z_{i + 2} - z)(z_{i + 3} - z)]}{[(z_{i} - z_{i + 1})(z_{i + 2} - z_{i + 1})(z_{i + 3} - z_{i + 1})]}\quad;\nonumber\\
A_2 &=& \frac{[(z_{i} - z)(z_{i + 1} - z)(z_{i + 3} - z)]}{[(z_{i} - z_{i + 2})(z_{i + 1} - z_{i + 2})(z_{i + 3} - z_{i + 2})]}\quad;\nonumber\\
A_3 &=& \frac{[(z_{i} - z)(z_{i + 1} - z)(z_{i + 2} - z)]}{[(z_{i} - z_{i + 3})(z_{i + 1} - z_{i + 3})(z_{i + 2} - z_{i + 3})]}\quad.\nonumber\\
\end{eqnarray}
The results obtained by the cubic interpolation method are almost similar to the results obtained by the linear interpolation method. It is important to emphasize again that the calibration results are completely independent of cosmological models used (for further discussion, see \cite {liang}).

In order to compare the GCGM with the observational data, the first step is to compute the theoretical luminosity distance $\mu$, 
\begin{equation}
\label{dl}
\mu^{th}=5\log \left( \frac{d_{L}}{M\!pc}\right) +25\quad ,
\end{equation}
with the relations for the GCGM described above. Here, as in \cite{liang2010, liang11}, by using only linear interpolating we have the 27 GRBs at $z \leq 1.4$ from the Union SNe Ia data and the 42 GRBs at $z > 1.4$ obtained with the five relations $(\tau_{lag}-L, V-L, L-E_p, E_{\gamma}-E_p, \tau_{RT}-L)$ calibrated with the sample at $z \leq 1.4$ that uses also the linear interpolation method. It is assumed that the GRBs luminosity relations do not evolve with redshift, so we could get the luminosity $(L)$ or energy $(E_{\gamma})$ of each burst at high redshift $(z > 1.4)$. The weighted average distance modulus from the five relations for each GRB is $\mu = (\sum_i \mu_i /\sigma^2_{\mu_i} )/(\sum_i \sigma^{-2}_{\mu_i})$, with its uncertainty $\mu_i = (\sum_i \sigma^{-2}_{\mu_i})^{-1/2}$, where the summations run from $1$ to $5$ over the five relations described above.

Considering a set of free parameters $\left\{{\bf p}\right\}$ the agreement between theory and observation is measured by minimizing the quantity, 
\begin{equation}
\chi^{2}\left({\bf p}\right)=\sum_{i=1}^{42}\frac{\left[\mu^{th}_{i}({\bf p}) - \mu^{obs}_{i}({\bf p})\right]^{2}}{\sigma_{i}^{2}},
\end{equation}
where $\mu^{th}$ and $\mu^{obs}$ are the theoretical value and the observed value of the luminosity distance for our model, respectively, and $\sigma$ means the error for each data point. We use Bayesian analysis to obtain the parameters estimations through the probability distribution function (PDF)
\begin{equation}
P = \mathcal{B} \,e^{-\frac{\chi^{2}(\bf{p})}{2}},
\end{equation}
where $\mathcal{B}$ is a normalization constant. A full Bayesian analysis is made by considering all free parameters of the model. However, we will study some particular Chaplygin configurations before a detailed analysis with 5 free parameters. With this strategy we hope to gain some intuition about the GRB data from the partial outcomes. Below, we will describe different Chaplygin-based cosmologies investigated in the present work. We show our results in Table 1-2 and in Figures 1-7. 

\begin{figure}[!h]
\includegraphics[width=0.32\textwidth]{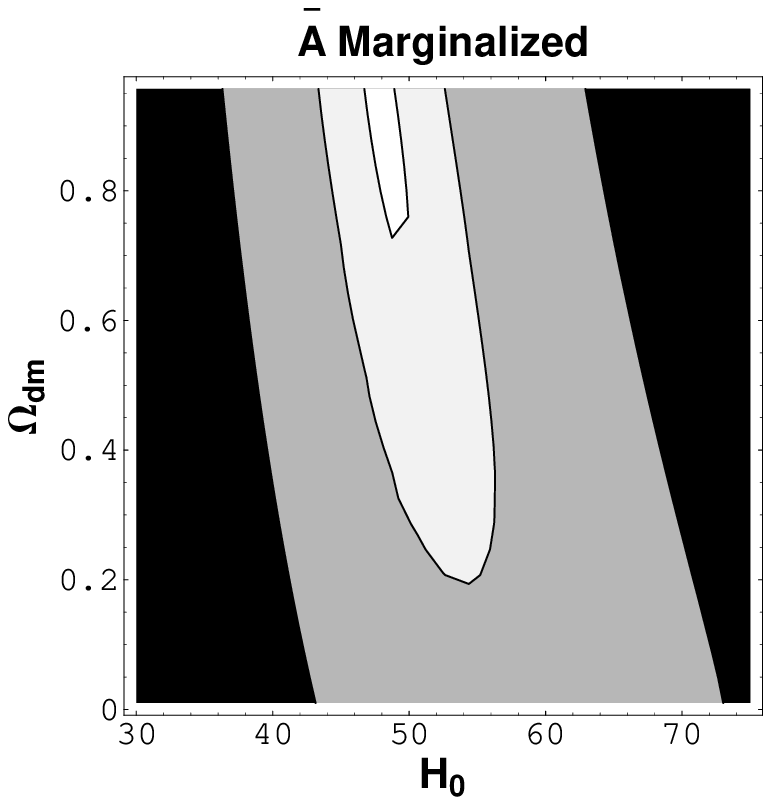}
\includegraphics[width=0.32\textwidth]{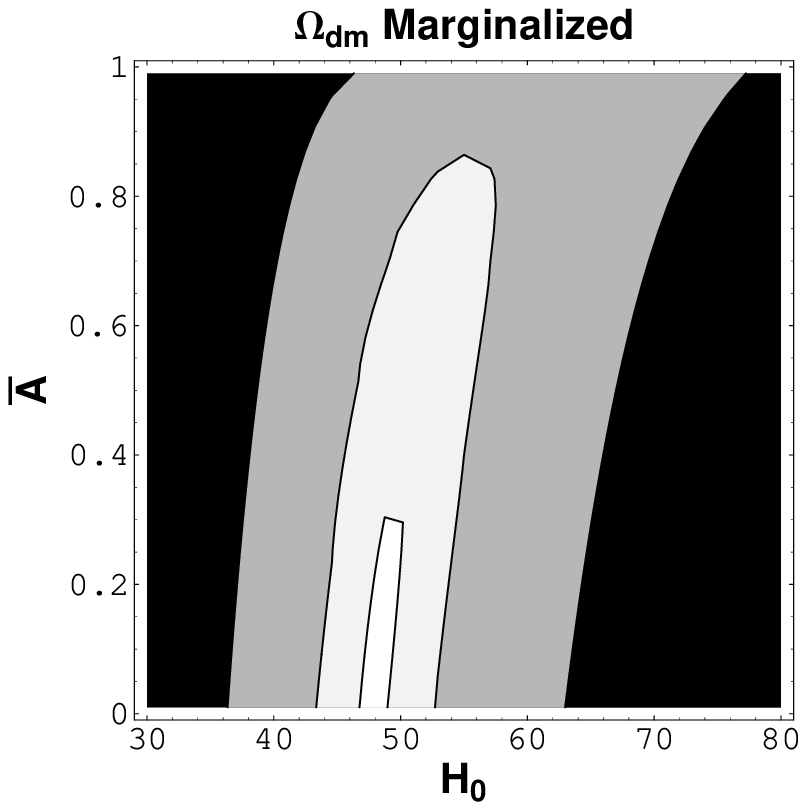}
\includegraphics[width=0.32\textwidth]{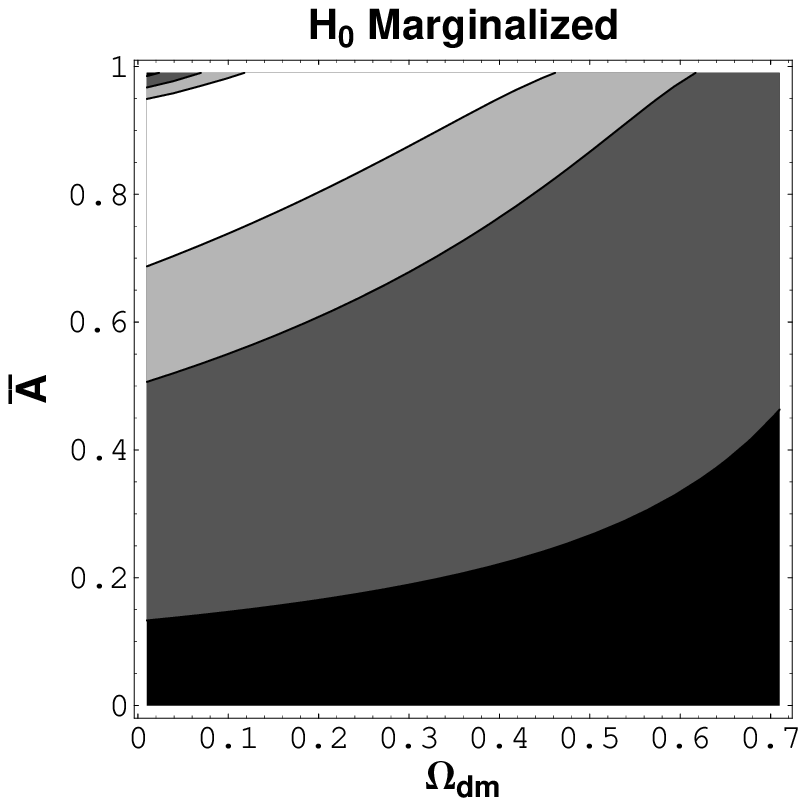}
\caption{Two-dimensional probability distribution function (PDF) for the free parameters in the CGM. The curves represent 99.73$\%$, 95.45$\%$ and 68.27$\%$ contours of maximum likelihood. The darker the region, the smaller the probability.}
\label{GRBc1}	
\end{figure}

\begin{figure}[!h]
\includegraphics[width=0.32\textwidth]{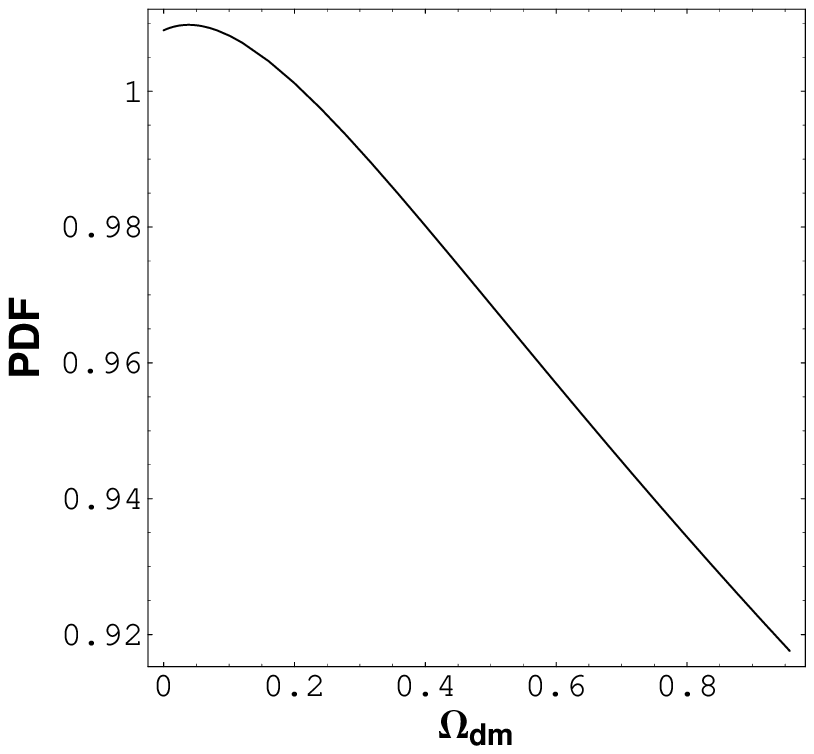}
\includegraphics[width=0.32\textwidth]{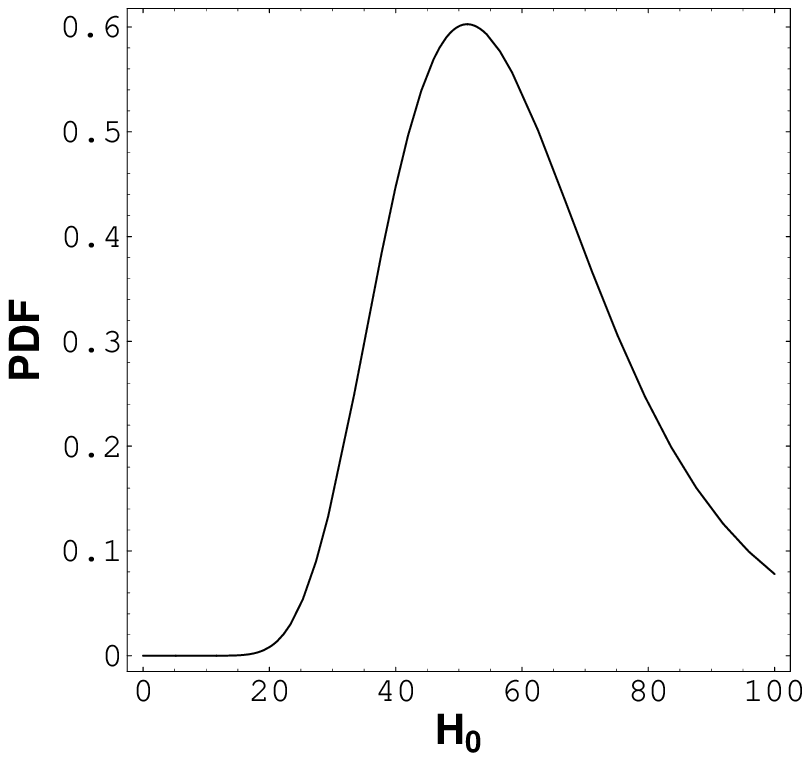}
\includegraphics[width=0.32\textwidth]{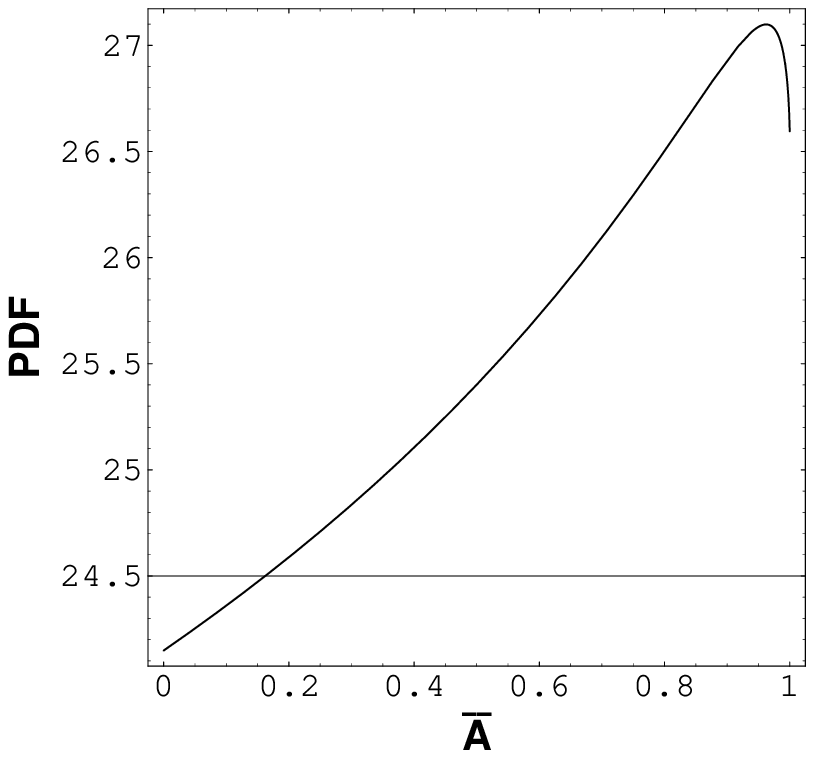}
\caption{One-dimensional PDFs for the three free parameters of the CGM.}
\label{GRBc2}	
\end{figure}

Our first step is to study the Chaplygin gas ($\alpha=1$). We remenber that this equation of state, as cited above, has also raised interest in particle physics thanks to its connection with string theory and its supersymmetric extension \cite{string}. We shall consider the prior information: $0 \leq \bar{A} \leq 1$, $0 \leq \Omega_{dm} \leq 0.957$ and $0 \leq H_{0} \leq 100$. The curves in Fig. \ref{GRBc1} represent the 99.73$\%$, 95.45$\%$ and 68.27$\%$ contours of maximum likelihood after the first marginalization, i.e, integration of the likelihood function over the non requerid parameter. However, the most robust parameter estimation is the central value for the parameter obtained in the maximum of the one-dimensional PDF as in Fig. \ref{GRBc2}.
From Fig. \ref{GRBc2} we can obtain the final parameter estimation. We found $\Omega_{dm}=0.04^{+0.59}_{-0.04}, \bar{A}=0.96^{+0.04}_{-0.61}$ and $H_{0}=51.3^{+9.5}_{-5.8}$ at 1$\sigma$ level. However, the dispersion in the GRB data is quite high. We compute these same estimatives using the Supernovae Constitution sample \cite{constitution} in order to compare the dispersion of these two samples. For the SN we found $\Omega_{dm}=0.00^{+0.40}_{-0.00}, \bar{A}=0.99^{+0.01}_{-0.41}$ and $H_{0}=59.7^{+2.1}_{-1.5}$ at 1$\sigma$ level. Some constraints on the Generalized Chaplygin gas have been placed using the Constitution data set \cite{Xu}. This allows a comparison between some of our results and the ones from Supernovae. In general, GRBs recover the results from SNe but with a high dispersion.

In our next analysis we relax the prior information about the Hubble parameter and leave it free to vary. We show the two-dimensional PDFs in the Fig. (\ref{GRB3}). In Fig. (\ref{GRB4}) the solid lines are the corresponding one-dimensional probabilities.

The above choice for the priors in the parameter $\alpha$ is conservative. With this choice we want to avoid a super luminal propagation in the sound speed formula. However, as argued in \cite{staro} the formula $v^{2}_{s}=\alpha \bar{A}$ represents the group sound velocity. Actually, in order to violate causality the wavefront velocity should exceed $1$ \cite{bril}. Considering this possibility we assume now $\alpha\geq0$ and compute the one-dimensional PDF as showed in dashed lines in Fig. \ref{GRB4}.

Until now we have considered a flat Universe in our analysis. In order to have a more general statistical analysis we allow a non-vanishing curvature in our model. The complete five-dimensional analysis is computationally hard but still feasible. We assume as prior information that our Universe deviates slightly from the flat model assuming $\Omega_{k}$ to vary between [-0.6,0.6]. For this case, we show the results in Fig. \ref{5pm}.

\begin{table}[h]
\begin{tabular}{|c|c|c|c|c|c|}\hline
   $ \mbox{Case} $ & $\alpha$ & $\bar{A}$  & $\Omega_{dm0}$  & $H_0$ & $\Omega_{k0}$  \\ \hline      
   CGM~$(\alpha=1) \rightarrow\mbox{Fig. 2} $   &   1  & $0.96^{+0.04}_{-0.61}$  & $0.04^{+0.59}_{-0.04}$  & $51.3^{+9.5}_{-5.8}$ & $0$  \\ \hline      
   GCGM~$(h=0.72)\rightarrow\mbox{Fig. 4}$ & $<<0$ & $0.98^{+0.02}_{-0.59}$ & $0.01^{+0.56}_{-0.01}$  & $72.0$  & $0$  \\ \hline       
   GCGM~$(0\leq\alpha\leq1)\rightarrow\mbox{Fig. 6} $  & $-4.3^{+4.8}_{-15.2}$   & $0.88^{+0.12}_{-0.54}$    & $0.10^{+0.52}_{-0.10}$  & $51.9^{+9.8}_{-5.6}$  & 0 \\ \hline      
   GCGM~$(\alpha\geq0)\rightarrow\mbox{Fig. 6} $  & $-4.3^{+4.8}_{-15.2}$   & $1.00^{+0.0}_{-0.34}$     & $0.00^{+0.61}_{-0.00}$   & $48.2^{+9.2}_{-5.3}$  & 0 \\ \hline      
   GCGM $\,\,\Omega_k \neq0~(0\leq\alpha<1)\rightarrow\mbox{Fig. 7} $  & $1.2^{+5.9}_{-7.4}$   & $0.64^{+0.24}_{-0.25}$     & $0.31^{+0.44}_{-0.20}$  & $56.2^{+10.1}_{-6.5}$ &  $-0.26^{+0.25}_{-0.26}$ \\ \hline
   GCGM $\,\,\Omega_k\neq0~(\alpha\geq0)\rightarrow\mbox{Fig. 7}$  & $1.2^{+5.6}_{-7.3}$  & $1.00^{+0.00}_{0.61}$ & $0.00^{+0.51}_{-0.00}$  & $52.3^{+8.9}_{-6.0}$ & $-0.53^{+0.29}_{-0.28}$ \\ \hline
      \end{tabular}
\caption{For the different Chaplygin-based cosmologies in the first column we show the final 1D estimation for the free parameters. The errors are computed at $1\sigma$ level.}
\end{table}
\begin{center}
\begin{table}[h]
\begin{tabular}{|c|c|c|c|c|c|c|}
   \hline
   $ \mbox{Figure} $ & $\Omega_{dm} \times H_0$ & $\bar{A} \times H_0$  & $\bar{A} \times \Omega_{dm0}$  & $\bar{A} \times \alpha$ & $\Omega_{dm} \times \alpha$ & $H_0 \times \alpha$  \\ \hline      
   1   &   (1.0 , 47.9)  & (0,48.1)  & (0.86 , 0.27)  & - & -  & -\\ \hline     
  3 & - & - & (0.86 , 0.27)  & (0.86 , 0.25) & (0.23 , -20) & - \\ \hline       
  5 & (49.8 , 1.0) & (0 , 48.1) & (0.86 , 0.26) & (0.86, 0.2)  & (0.23 , -12.0) & (50, +10.0)  \\ \hline      
  
      \end{tabular}
      
\caption{Best fit values for the 2D PDFs.}
\end{table}
\end{center}


\begin{figure}[!h]
\includegraphics[width=0.32\textwidth]{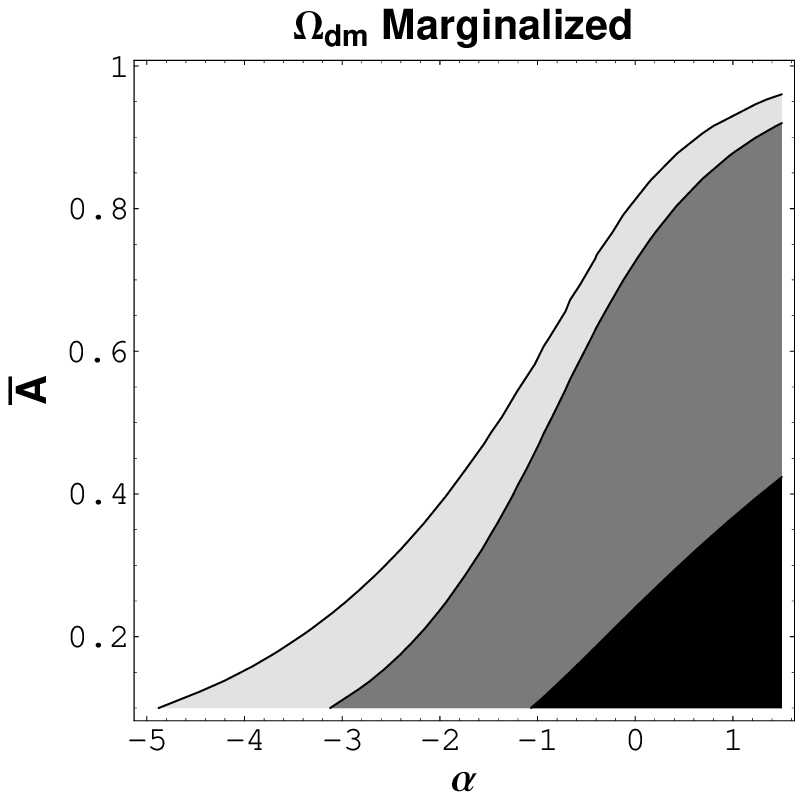}
\includegraphics[width=0.32\textwidth]{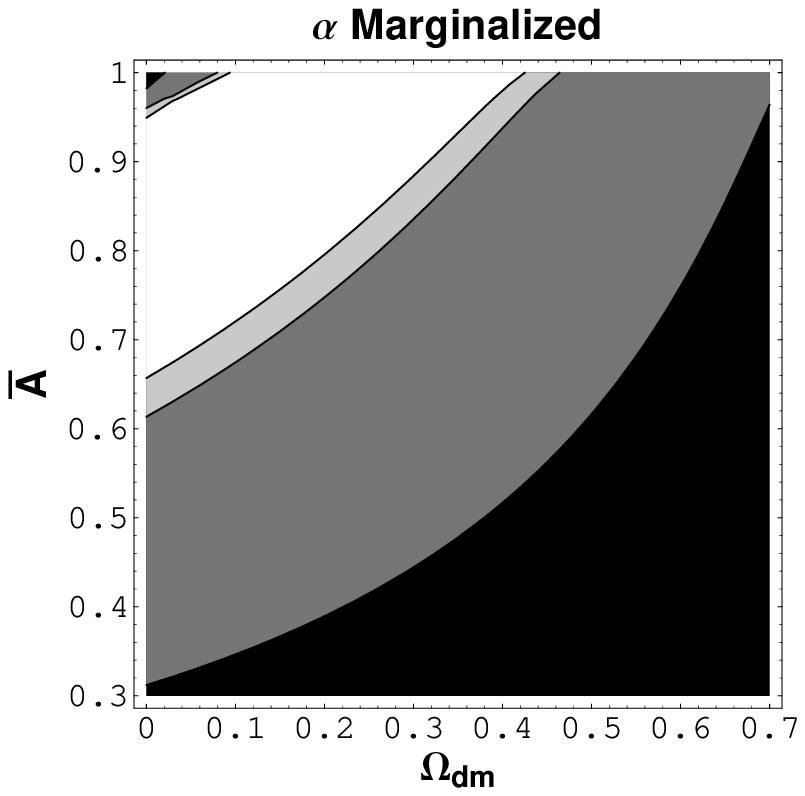}
\includegraphics[width=0.32\textwidth]{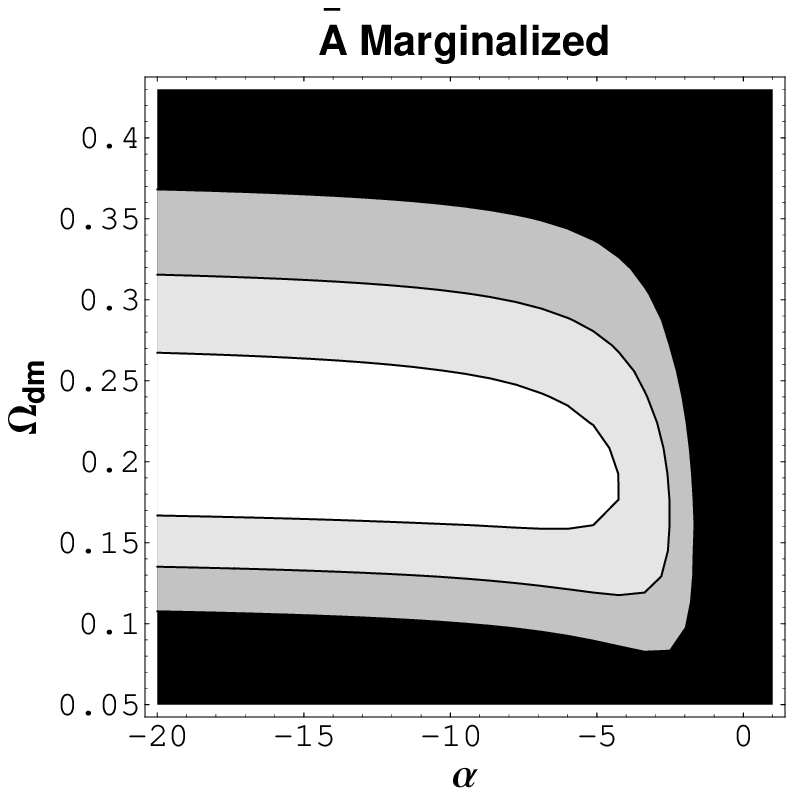}
\caption{Two-dimensional PDFs for the GCGM fixing $H_{0}=72~km~s^{-1} ~Mpc^{-1}$. The curves represent 99.73$\%$, 95.45$\%$ and 68.27$\%$ contours of maximum likelihood. The darker the region, the smaller the probability.}
\label{GRB1}	
\end{figure}
\begin{figure}[!h]
\includegraphics[width=0.32\textwidth]{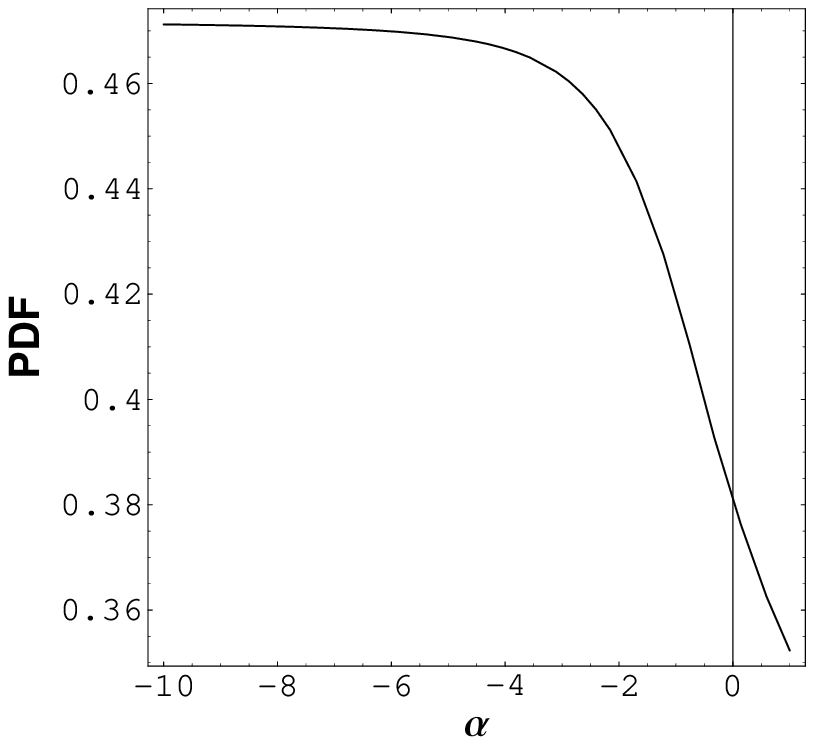}
\includegraphics[width=0.32\textwidth]{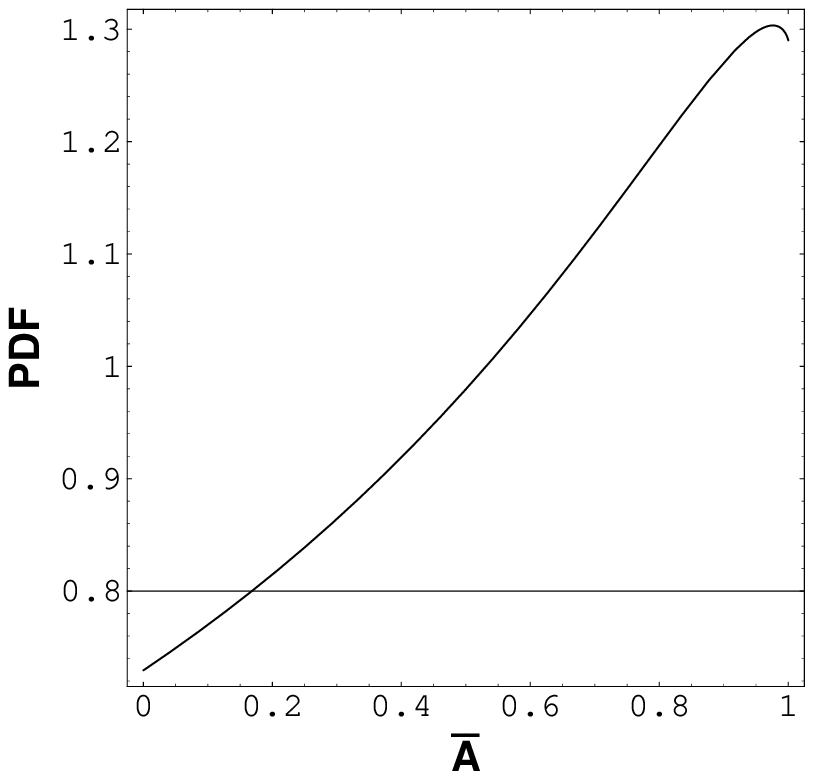}
\includegraphics[width=0.32\textwidth]{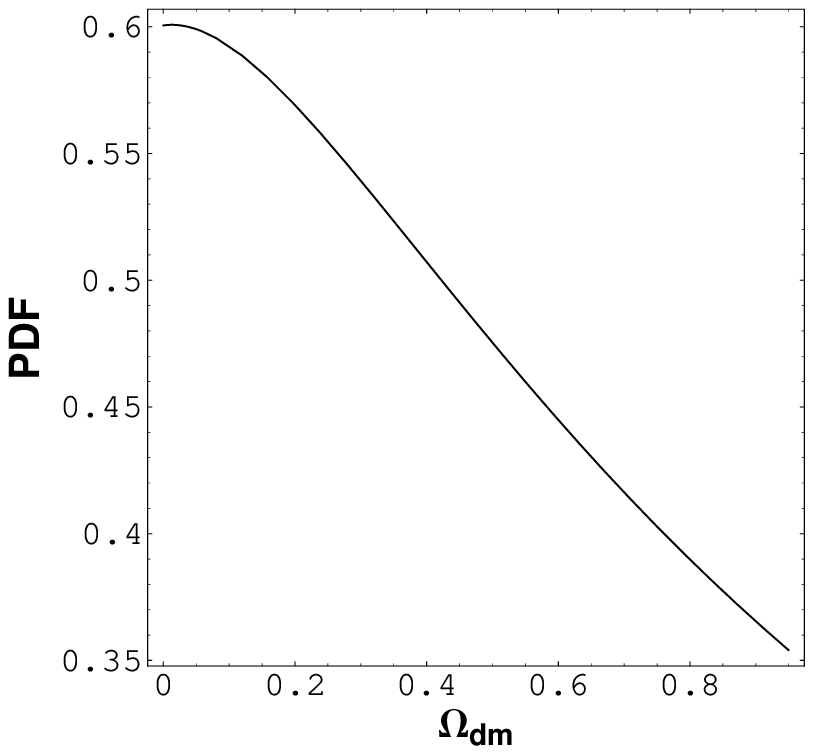}
\caption{One-dimensional PDFs for the three free parameters of the GCGM when $H_{0}=72~km~s^{-1} ~Mpc^{-1}$.}
\label{GRB2}	
\end{figure}

\begin{figure}[!t]
\includegraphics[width=0.32\textwidth]{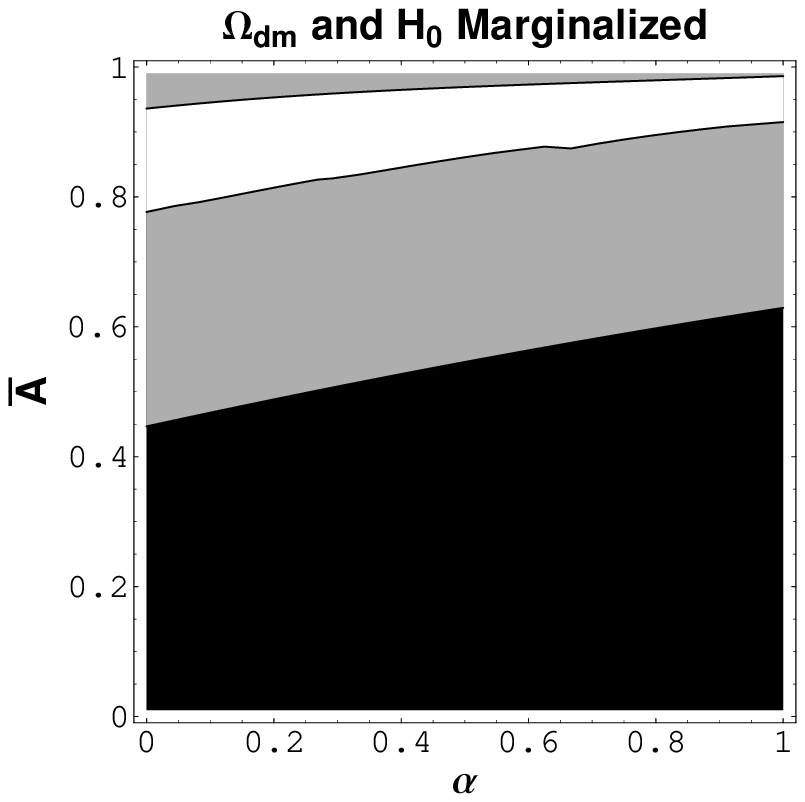}
\includegraphics[width=0.32\textwidth]{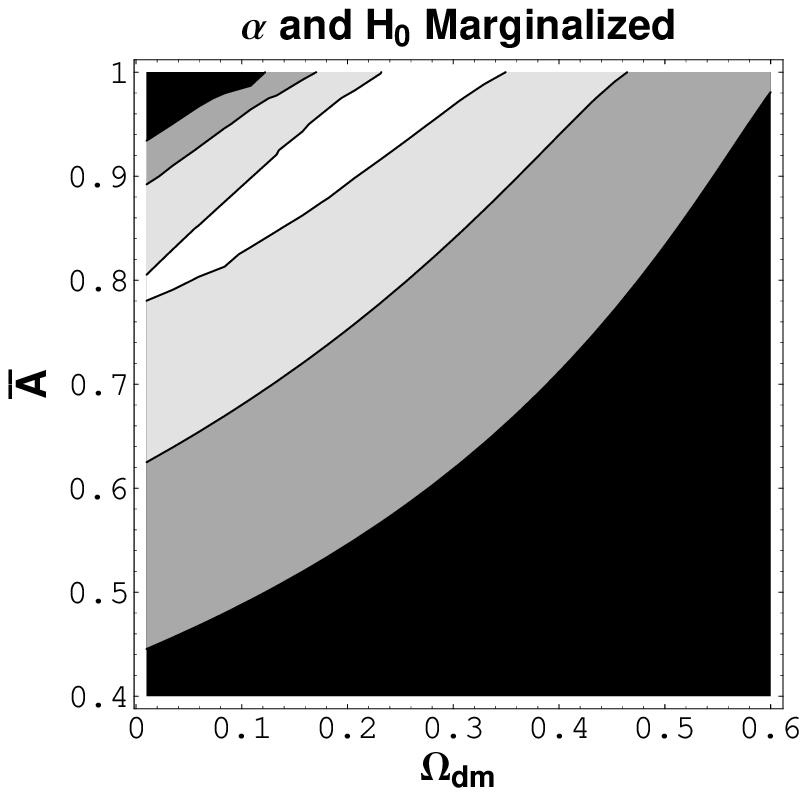}
\includegraphics[width=0.32\textwidth]{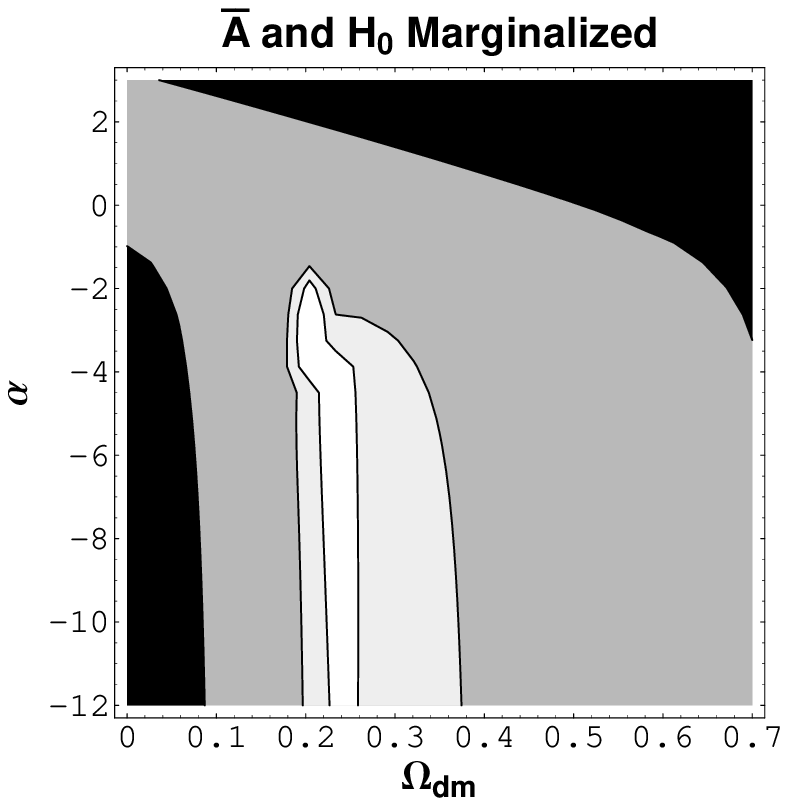}
\includegraphics[width=0.32\textwidth]{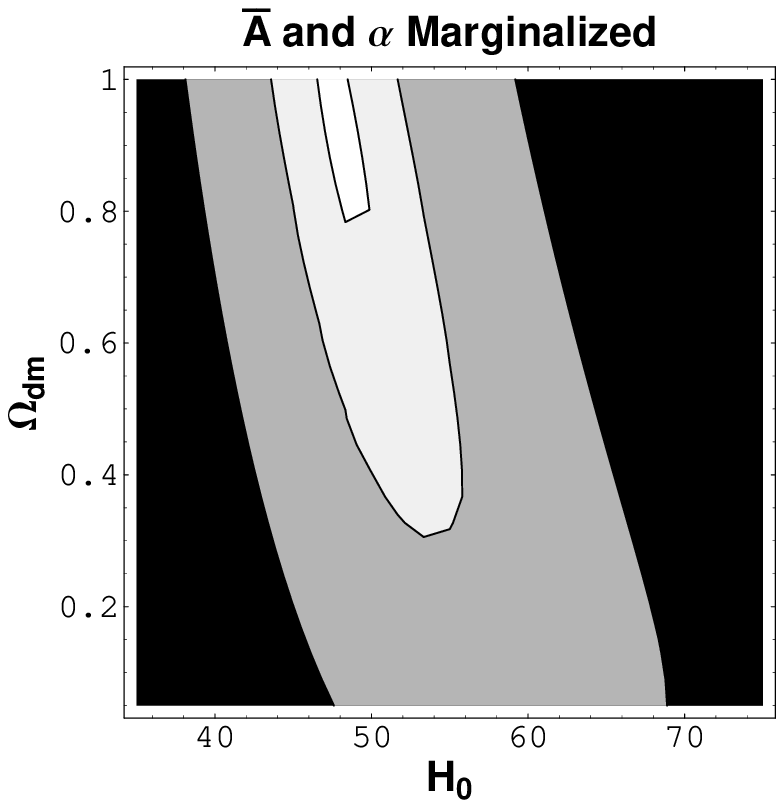}
\includegraphics[width=0.32\textwidth]{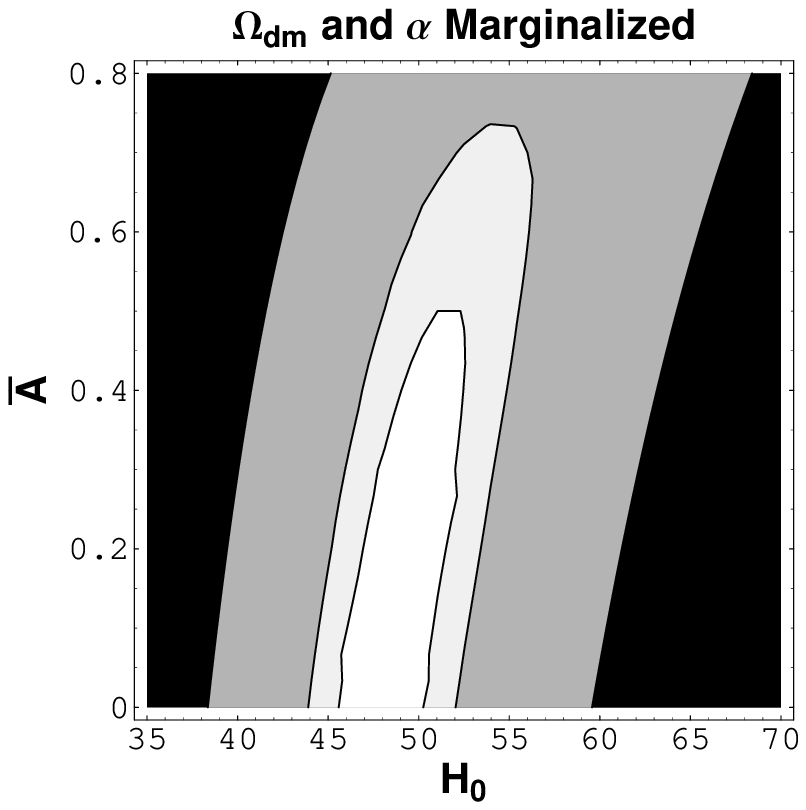}
\includegraphics[width=0.32\textwidth]{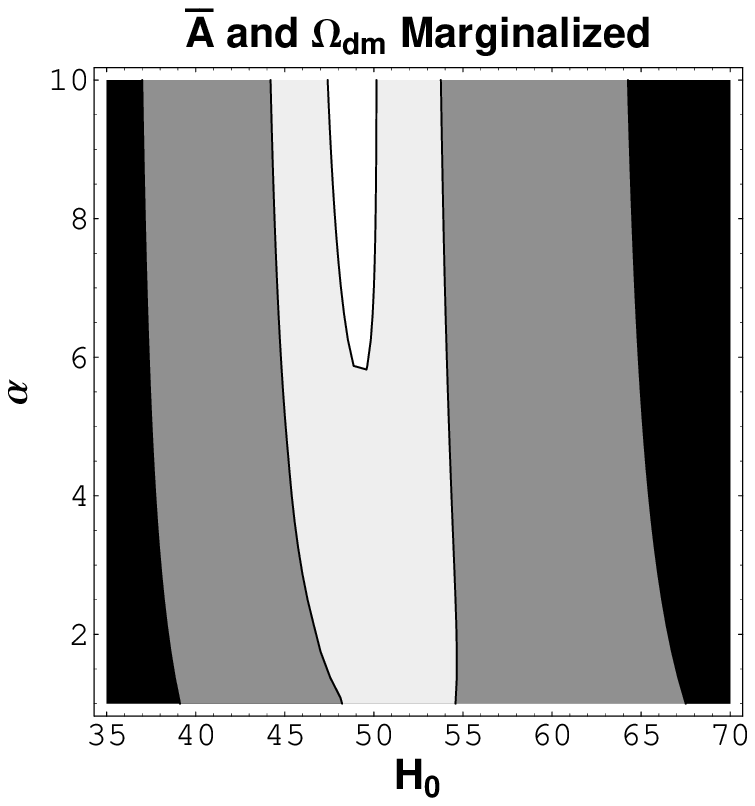}
\caption{The same as Fig. (1) but considering four free parameters for the GCGM and the prior $0\leq\alpha\leq1$.}
\label{GRB3}	
\end{figure}

\begin{figure}[!h]
\includegraphics[width=0.24\textwidth]{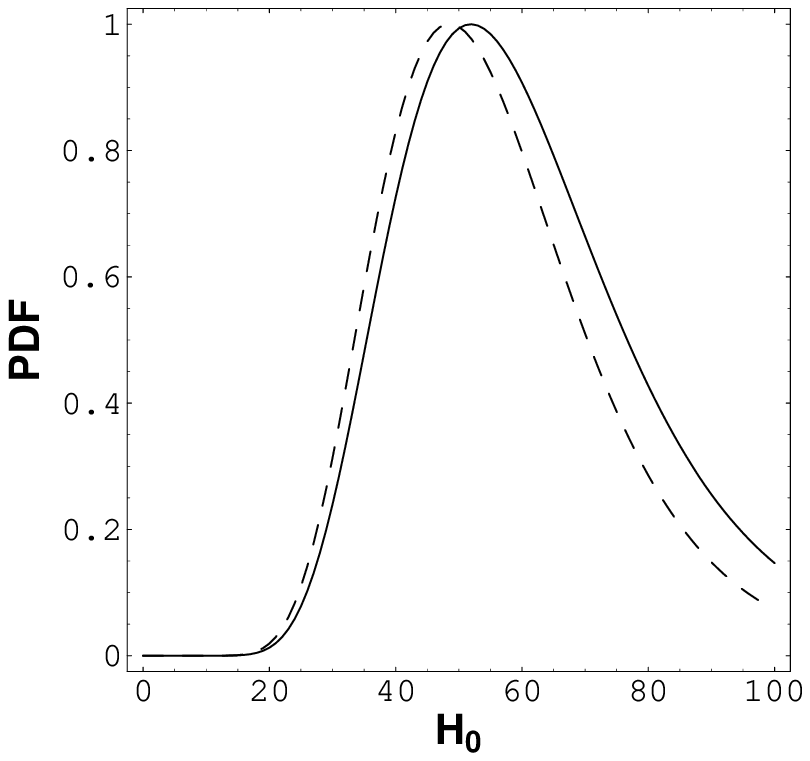}
\includegraphics[width=0.24\textwidth]{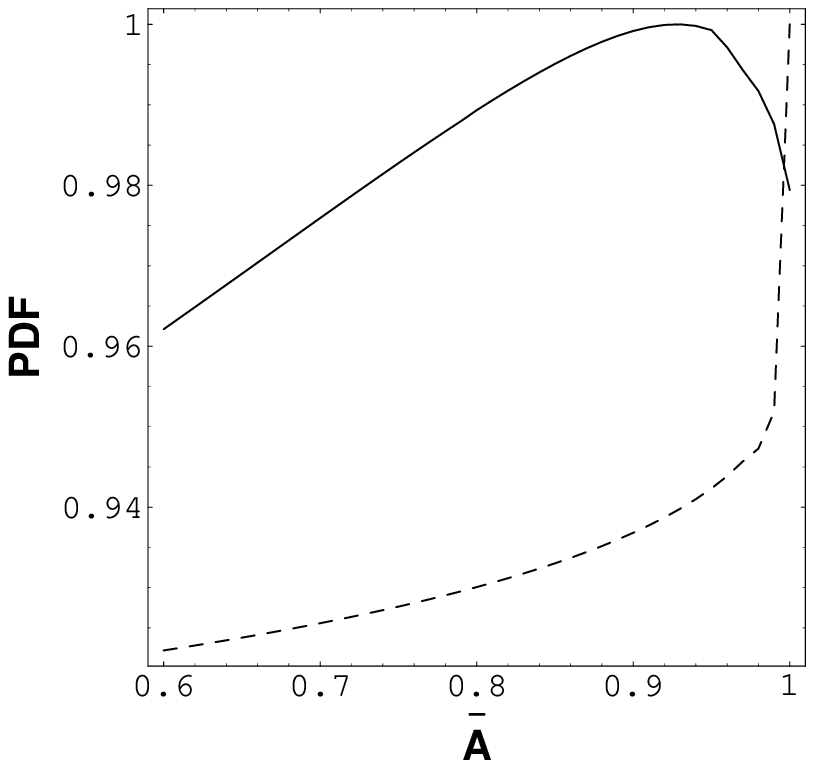}
\includegraphics[width=0.24\textwidth]{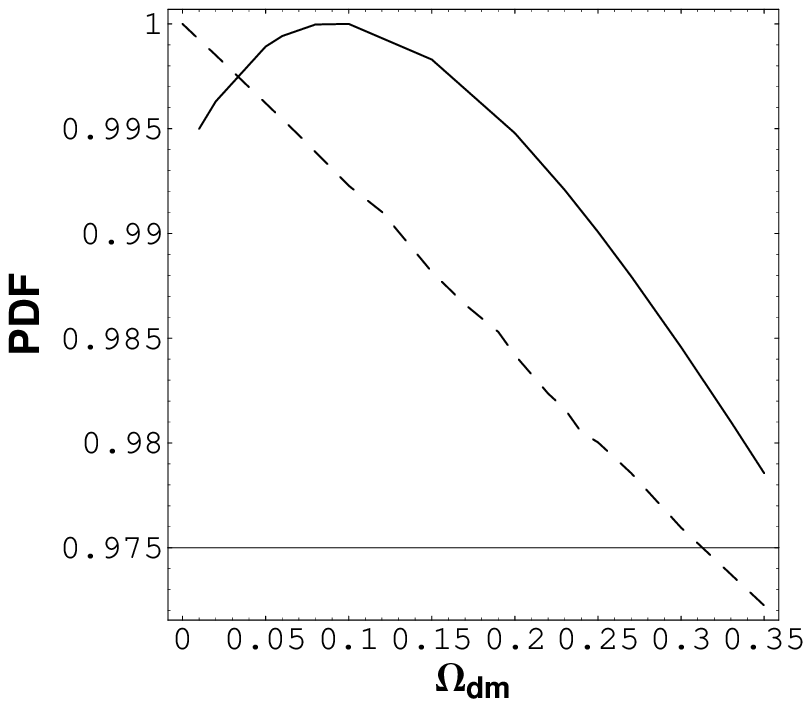}
\includegraphics[width=0.24\textwidth]{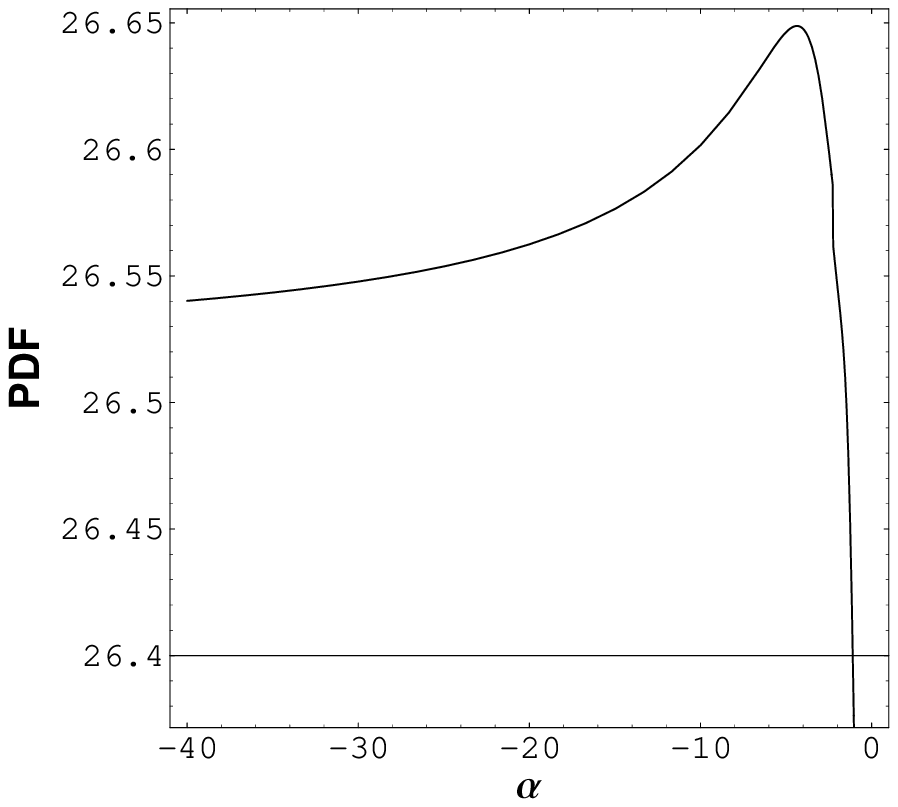}
\caption{One-dimensional PDF for the GCGM free parameters when $H_0$ is free to vary. The solid lines correspond to the prior choice $0\leq\alpha\leq1$ while dashed lines correspond to the prior $\alpha\geq 0$. The final estimation for the parameter $\alpha$ does not depend on its prior information.}
\label{GRB4}	
\end{figure}
       
\begin{figure}[!h]
\includegraphics[width=0.25\textwidth]{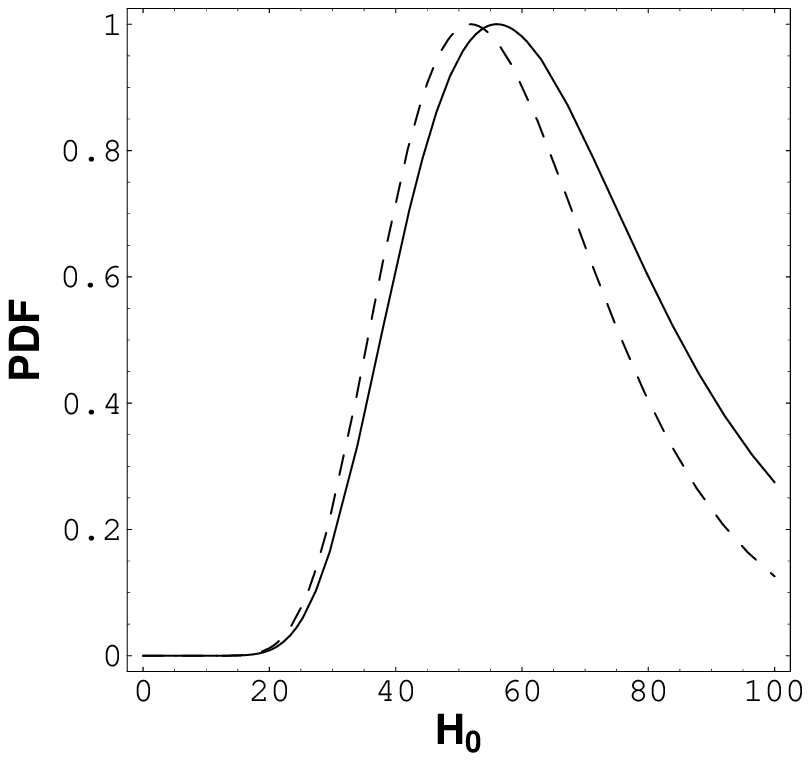}
\includegraphics[width=0.25\textwidth]{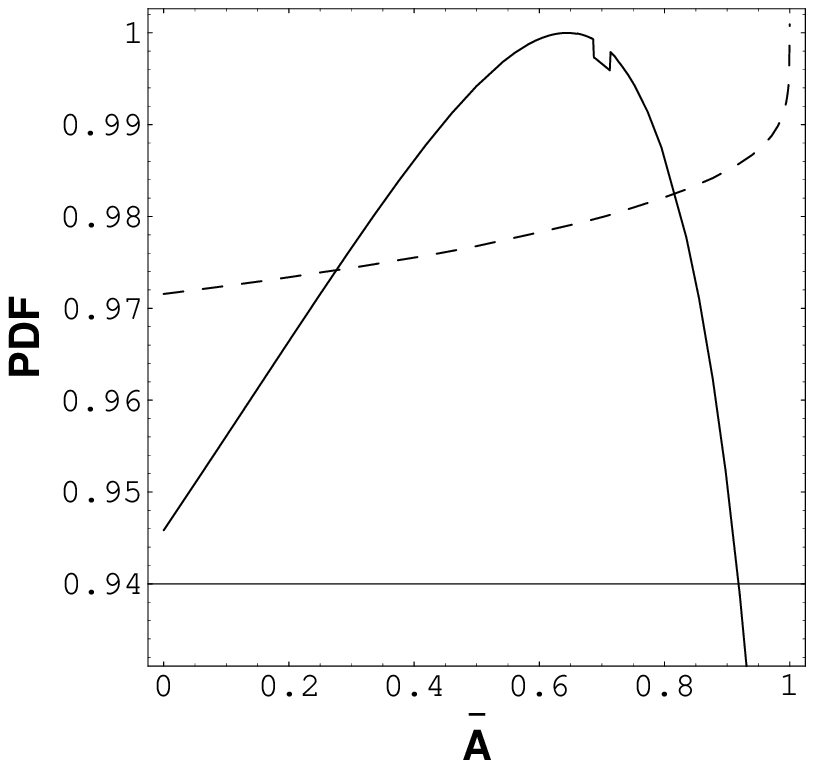}
\includegraphics[width=0.25\textwidth]{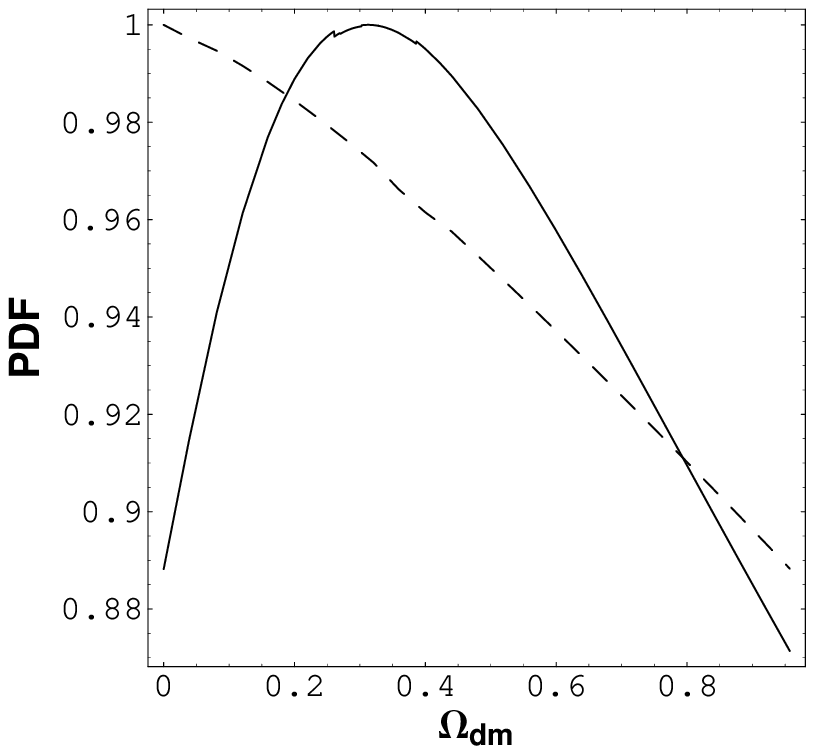}
\includegraphics[width=0.25\textwidth]{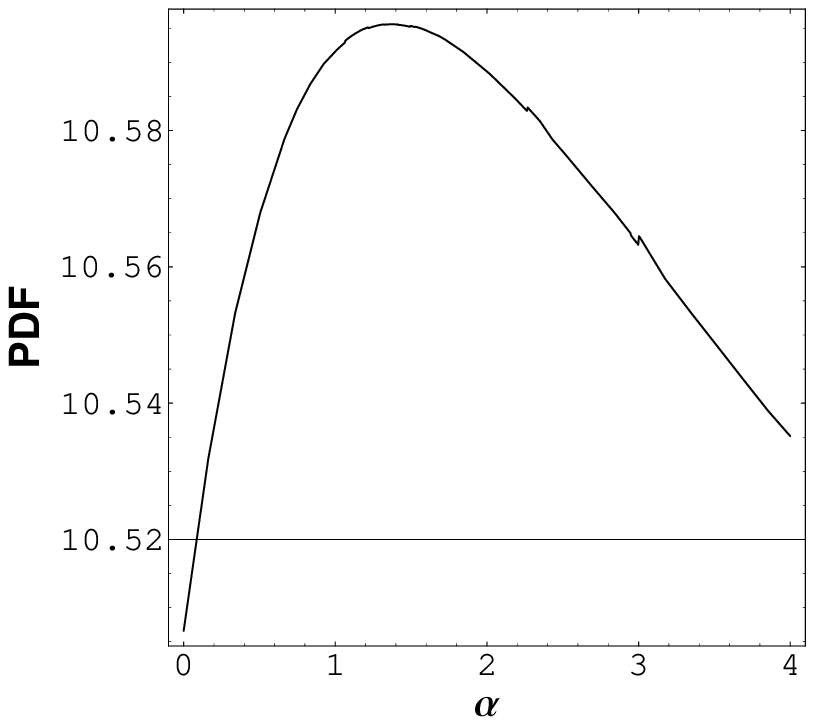}
\hspace{1.9cm}
\includegraphics[width=0.25\textwidth]{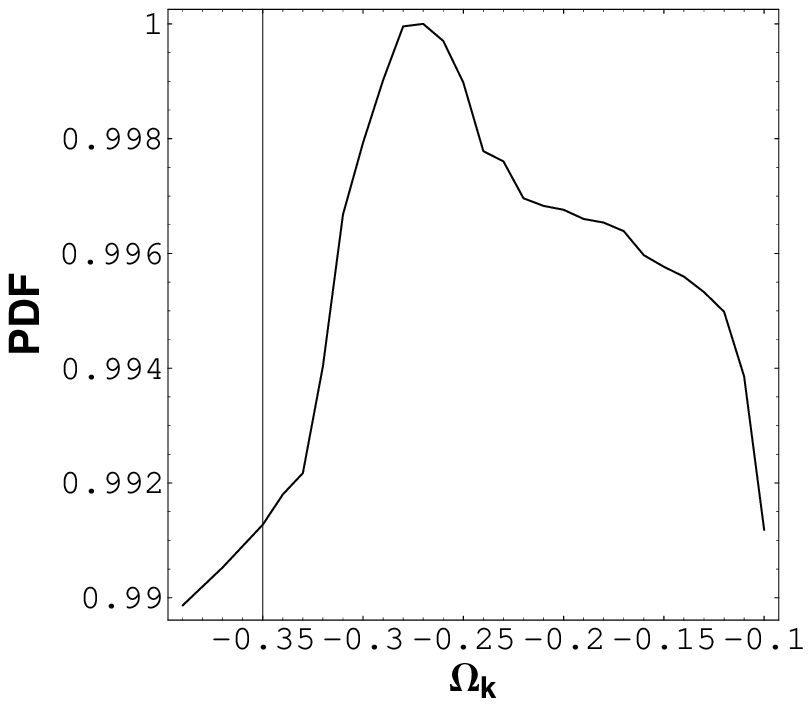}
\hspace{1.8cm}
\includegraphics[width=0.25\textwidth]{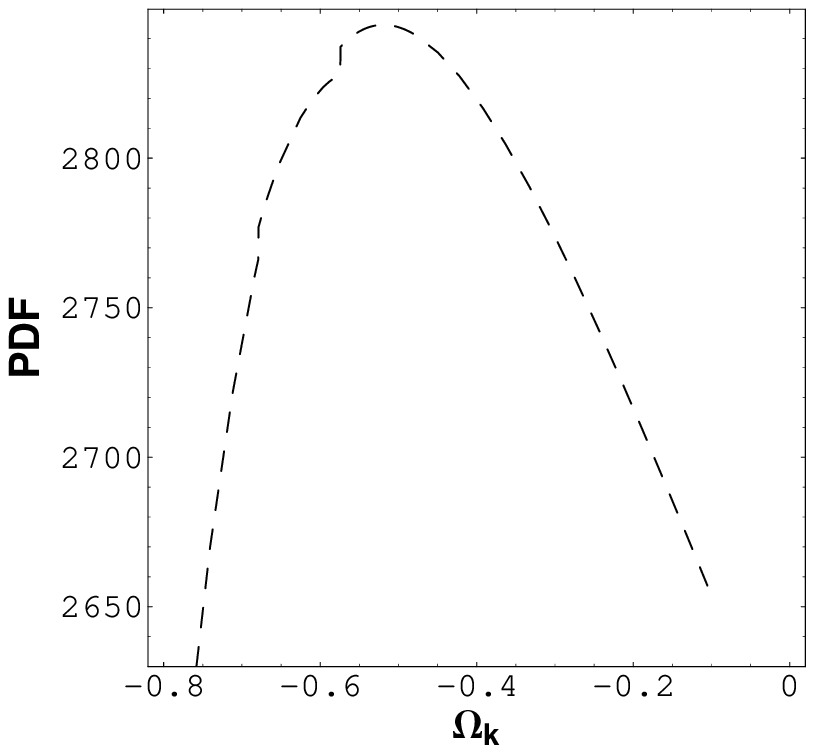}
\caption{One-dimensional PDF for the GCGM free parameters when we allow a non-vanishing curvature. The solid lines correspond to the prior choice $0\leq\alpha<1$ while dashed lines correspond to the prior $\alpha\geq0$. The final estimation for the parameter $\alpha$ does not depend on its prior information.}
\label{5pm}	
\end{figure}       

\section{Discussion and Conclusions}
\par

In this study we have analyzed the Chaplygin gas model with a sample of 42 GRBs. Although the use of GRBs as a cosmological tool is a promising way to probe cosmology at high redshifts we have verified that the available data is still insufficient to impose precise constraints in cosmological models. As observed in our analysis, the dispersion is still high when compared with others observational data sets. However we hope that with the future data from the final Swift BAT Catalog we will be able to put strong constraints on the dark energy/matter properties.  
\par
In our analysis, the unification scenario was not imposed from the beginning. This means that we allow an extra dark matter contribution ($\Omega_{dm}$) in our calculations in order to probe whether the unification scenario is favoured. In our first analysis the free parameters ($\bar{A},\Omega_{dm}$ and $H_{0}$) of the Chaplygin gas ($\alpha=1$) were well constrained. Our results are in agreement with the Supernova results \cite{fabris03}. The only difference is that we find a lower value for the Hubble parameter, $H_{0}=51.3^{+9.2}_{-5.7}$ (1$\sigma$). However, it is possible to find in the literature similar results for the parameter $H_{0}$ \cite{H0}. 

In our second analysis, in order to check the behaviour of the model when $H_{0}=72~km~s^{-1} Mpc^{-1}$ we leave $\alpha$ free, that is the so called Generalized Chaplygin Gas Model. From Figs. \ref{GRB1} and \ref{GRB2} the unification scenario is again favoured. However, the uncertainties are still high. The parameter $\alpha$ assumes a large negative value. There is no any peak in the parameter $\alpha$ distribution and the probability remains constant for negative values. For the background dynamics the region ($\alpha<-1$) represents a behavior different from the matter dominated phase when structures start to form. On the other hand, negative values for $\alpha$ imply an imaginary sound velocity, leading to small scale instabilities at the perturbative level. Rigourously, the general situation is more complex: such instabilities for fluids with negative pressure may disappear if the hydrodynamical approach is replaced by a more fundamental description using, e.g., scalar fields. However, this is not true for the Chaplygin gas: even in a fundamental approach, using for example the Born-Infeld action, the sound speed may be negative if $\alpha<0$. Perhaps the restriction $\alpha\geq0$ must be imposed for all observational tests.

We work also with a set of four free parameters. Varying all four parameters, the preceding results are confirmed. Leaving the parameter $H_{0}$ free to vary, we confirm that the hypersurface $H_{0}=72~km~s^{-1} ~Mpc^{-1}$ doesn't represent the maximum probability in the 4-D parameters phase space. Chaplygin gas models show values lower than $H_{0}=72~km~s^{-1} Mpc^{-1}$ \cite{fabris03}. 
We observe also that there is a significant difference in the final parameter estimation when we consider the prior $\alpha\geq0$, instead of $0\leq\alpha<1$. For instance, the unification scenario ($\Omega_{dm}= 0$) is favoured only with the choice $\alpha\geq0$. Moreover, there is now a peak in the $\alpha$ distribution at $\alpha=-4.3^{+4.8}_{-15.2}$ but with a high dispersion. Again, negative values for $\alpha$ are favored, despite the two-dimensional PDF ($\alpha$ x $H_{0}$) in Fig.\ref{GRB3} indicates a high probability for $\alpha>6$. Such a contradiction seems to be an artifact of the marginalization process as can be seen also in the two-dimensional PDFs ($\Omega_{dm}$ x $H_{0}$) and ($\bar{A}$ x $H_{0}$) in Fig. \ref{GRB3}. These plots confirm that the final 1D estimation can be very different from the partial 2D ones. This diferrence is due to the integration of the probability function over the adopted prior values of the remmaning parameters.

The analysis with five free parameters confirm some of the previous results. Negative curvature is prefered as well as Sn data data \cite{fabris03}. Also, the parameter $\alpha$ is now estimated with a positive value, in constrast with the previous results.  

\begin{figure}[!h]
\begin{center}
\includegraphics[width=0.32\textwidth]{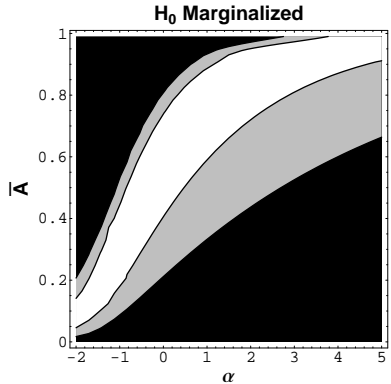}
\caption{Constrainst on the free parameters of the unified ($\Omega_{dm}=0$) GCG model.}
\label{UnifGRB}
\end{center}	
\end{figure}

The Chaplygin gas parameters have been estimated in many papers, considering different analysis and several observational data sets. Constraints critically depend on whether one treats the Chaplygin gas as true quartessence (replacing both dark matter and dark energy) or if one allows it to coexist with a normal dark matter component. The former situation is widely considered in the literature. As we leave the density parameter $\Omega_{dm}$ free to vary in all the cases analysed here, it is not possible to directly compare our results with unified Chaplygin cosmologies  unless we assume the prior $\Omega_{dm}=0$. This case has been studied using GRBs and other probes in reference \cite{liang11}. For a comparison with this reference, figure \ref{UnifGRB} shows the two-dimmensional probability for the free parameter of the unified ($\Omega_{dm}=0$) GCG model. The best fit occurs at ($\alpha=0.15, \bar{A}=0.75$). This result agrees (at 1$\sigma$) with the joint analysis showed in \cite{liang11}.

The analysis of section 3 can be compared with \cite{fabris03}, where the influence of a free $\Omega_{dm}$ parameter on the final estimations was taken into account. Our results have high confidence with the results obtained in \cite{fabris03}.

Finally, we remark that, perturbative analysis of Chaplygin models, for instance, reveals a large positive value ($\alpha>>200$) for the parameter $\alpha$ \cite{fabris04} while kinematic tests show values negatives or close to zero. At the background level, the crossing of different data sets (including for example SNe, CMB, BAO, H(z) data and galaxy cluster mass fraction) will provide a more accurate scenario for each Chaplygin-based cosmology studied in this work. We leave this analysis, including the perturbative study, for a future work.
   
\section*{Acknowledgements}

R.C.F., S.V.B.G. and H.E.S.V. thank DAAD (Germany) and CNPq, CAPES and FAPES (Brazil) for partial financial support. S.V.B.G. thanks the Laboratoire d'Annecy-le-Vieux de Physique Theorique (France) and H.E.S.V. thanks the Fakult\"at f\"ur Physik, Universit\"at Bielefeld (Germany) for kind hospitality during part of the elaboration of this work. We thank Winfried Zimdahl and Christian Byrnes for the comments and suggestions.

\end{document}